\documentclass[prb,preprint]{revtex4}

\newcommand{\gnr}{GNR}
\newcommand{\method}{NEGF method}

\usepackage{latexsym, amssymb, amsmath, revsymb}
\usepackage{graphicx, subfigure}

\begin{document}

\title{Perfect Reflection of Chiral Fermions in Gated Graphene Nanoribbons}

\author{J.M. Kinder}
	\email{jesse.kinder@gmail.com}
\author{J.J. Dorando}
\author{H. Wang}
\author{G.K.-L. Chan}
\affiliation{Department of Chemistry and Chemical Biology, Cornell University, Ithaca, New York 14850}

\begin{abstract}

We describe the results of a theoretical study of transport through gated metallic graphene nanoribbons using a non-equilibrium Green function method. Although analogies with quantum field theory predict perfect transmission of chiral fermions through gated regions in one dimension, we find \emph{perfect reflection} of chiral fermions in armchair ribbons for specific configurations of the gate. This effect should be measurable in narrow graphene constrictions gated by a charged carbon nanotube.

\end{abstract}

\maketitle

Graphene nanoribbons offer an opportunity to study the unusual transport properties of graphene in a confined geometry. The width and edge structure of a ribbon determine its electronic properties, and potentials that couple states in different bands can lead to effects that have no analog in graphene sheets. This theoretical transport study suggests one such effect: perfect reflection of chiral fermions.

Graphene, carbon nanotubes, and graphene nanoribbons (\gnr{}s) are derived from a honeycomb lattice, which has a two-atom basis. In the effective mass description of these systems, the amplitudes of an electron wave function on the two inequivalent sublattices is described by a pseudospin.\cite{castroneto2007epg} If the dispersion relation is linear, the low-energy properties can be described with an effective theory of massless fermions. These massless fermions are chiral: their pseudospin is parallel to their direction of motion, and electrons moving in opposite directions have opposite pseudospins.

Chiral fermions cannot be reflected by a scalar potential such as an electrostatic gate or a charged particle. This requires reversing the direction of motion \emph{and} flipping the pseudospin, but a scalar potential has no effect on the pseudospin. This property is responsible for the ``Klein paradox'' in graphene\cite{katsnelson2006cta} and the ``absence of backscattering'' in metallic carbon nanotubes.\cite{ando1998isc, mceuen1999dpb}

\gnr{}s with general edge structures have a nearly flat band of exponentially localized metallic edge states.\cite{nakada1996esg, akhmerov2008bcd} However, \gnr{}s with armchair edges can be metallic or semiconducting depending on their width. Metallic armchair ribbons have a linear dispersion relation, and their low-energy excitations are chiral fermions. Electrons in these ribbons should have long mean free paths and transport should be insensitive to disorder. Chiral fermions are specific to armchair ribbons.

\begin{figure}[htb]
	\begin{centering}
		\includegraphics[width=0.48\textwidth]{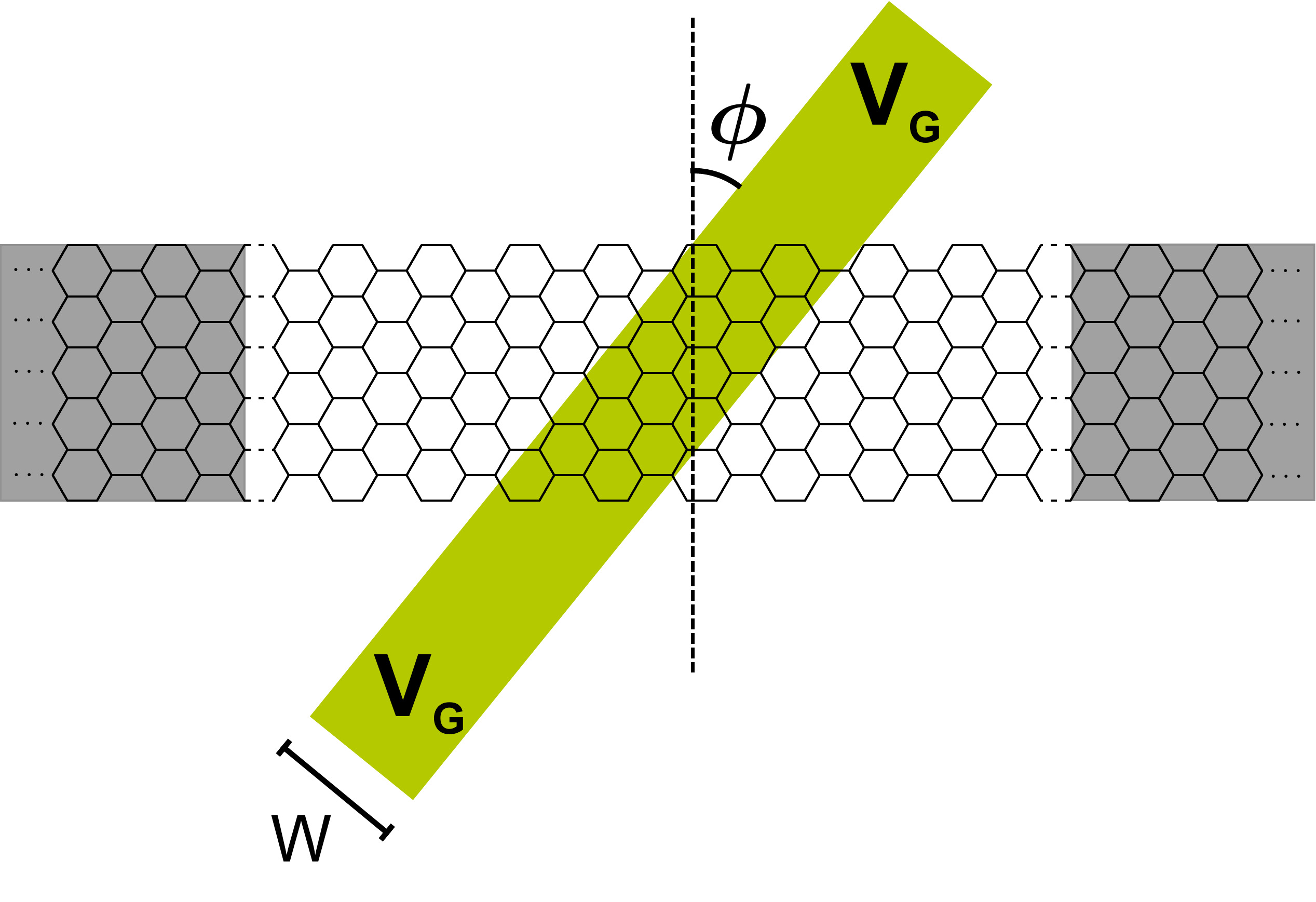}
		\caption{Schematic of a graphene nanoribbon with a rotated gate. The conductance of the device depends on the gate voltage, orientation, and width, as well as the width and edge structure of the ribbon. In calculations, the nanoribbon is divided into two ``leads'' (shaded) and a ``sample'' that includes the gated region.}
		\label{fig:device}
	\end{centering}
\end{figure}

\begin{figure}[hbt]
	\begin{center}
	\subfigure[Armchair Ribbon]{\includegraphics[width=0.32\textwidth]{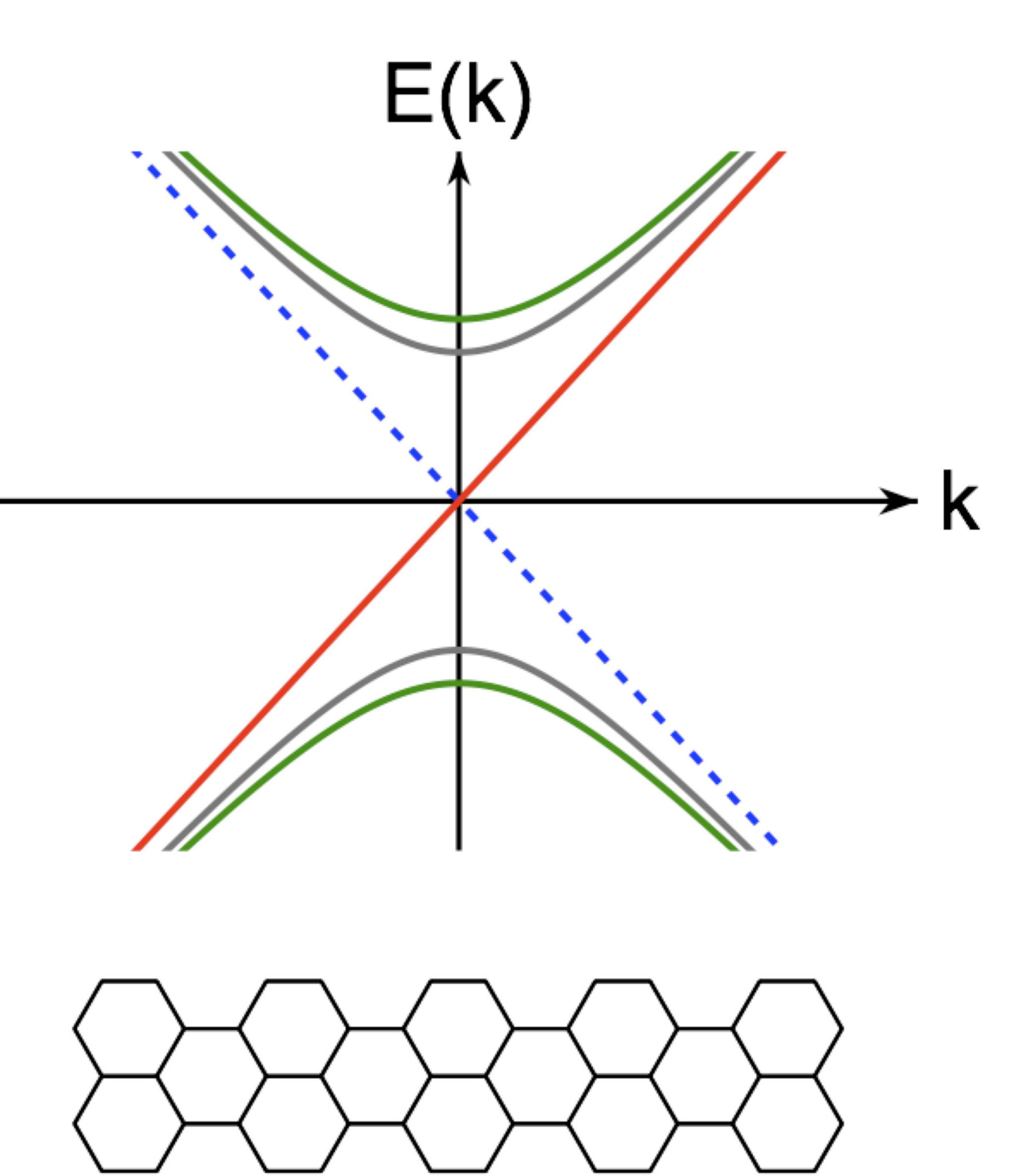}} \
	\subfigure[Zigzag Ribbon]{\includegraphics[width=0.32\textwidth]{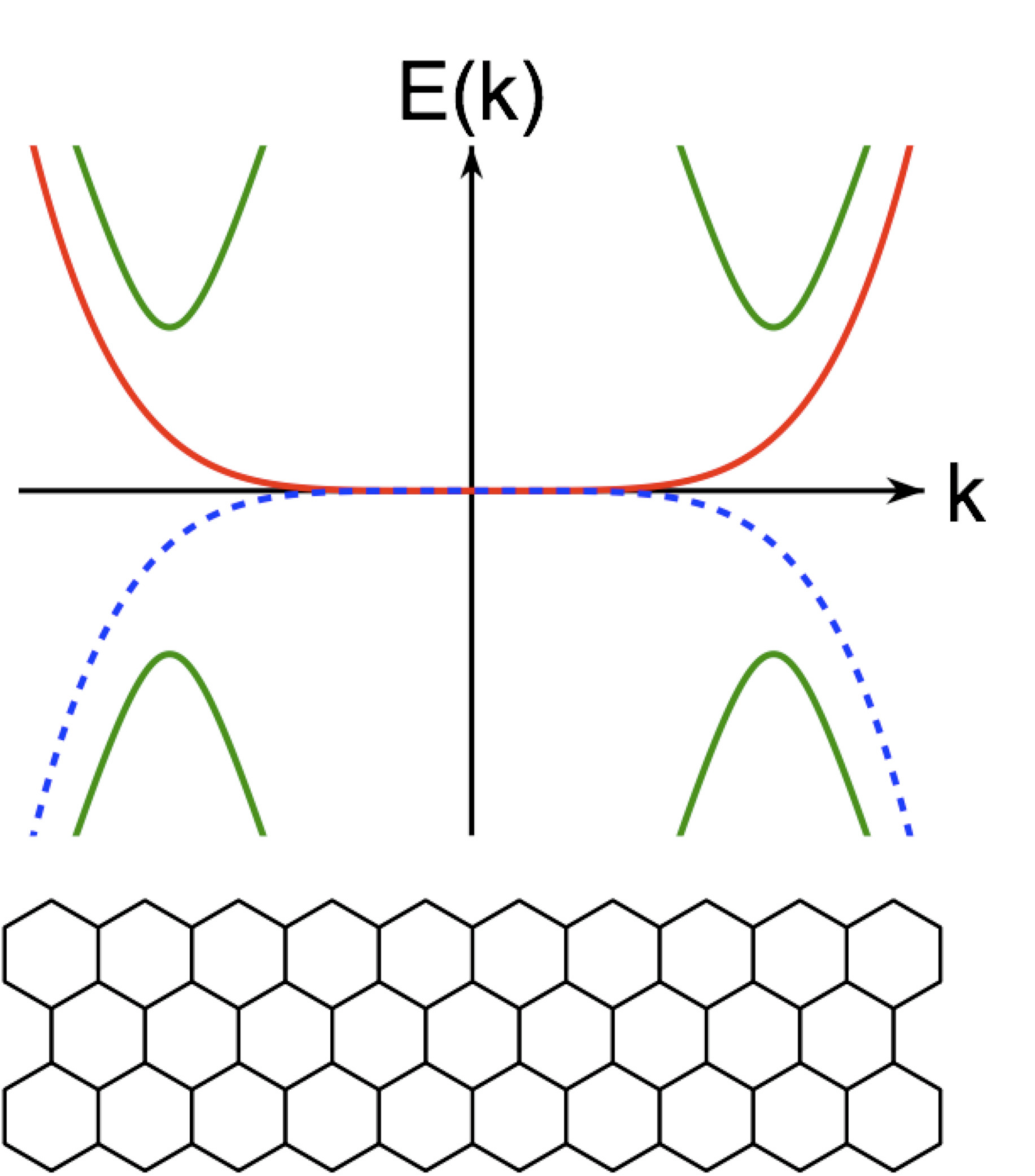}} \
	\subfigure[Anti-Zigzag Ribbon]{\includegraphics[width=0.32\textwidth]{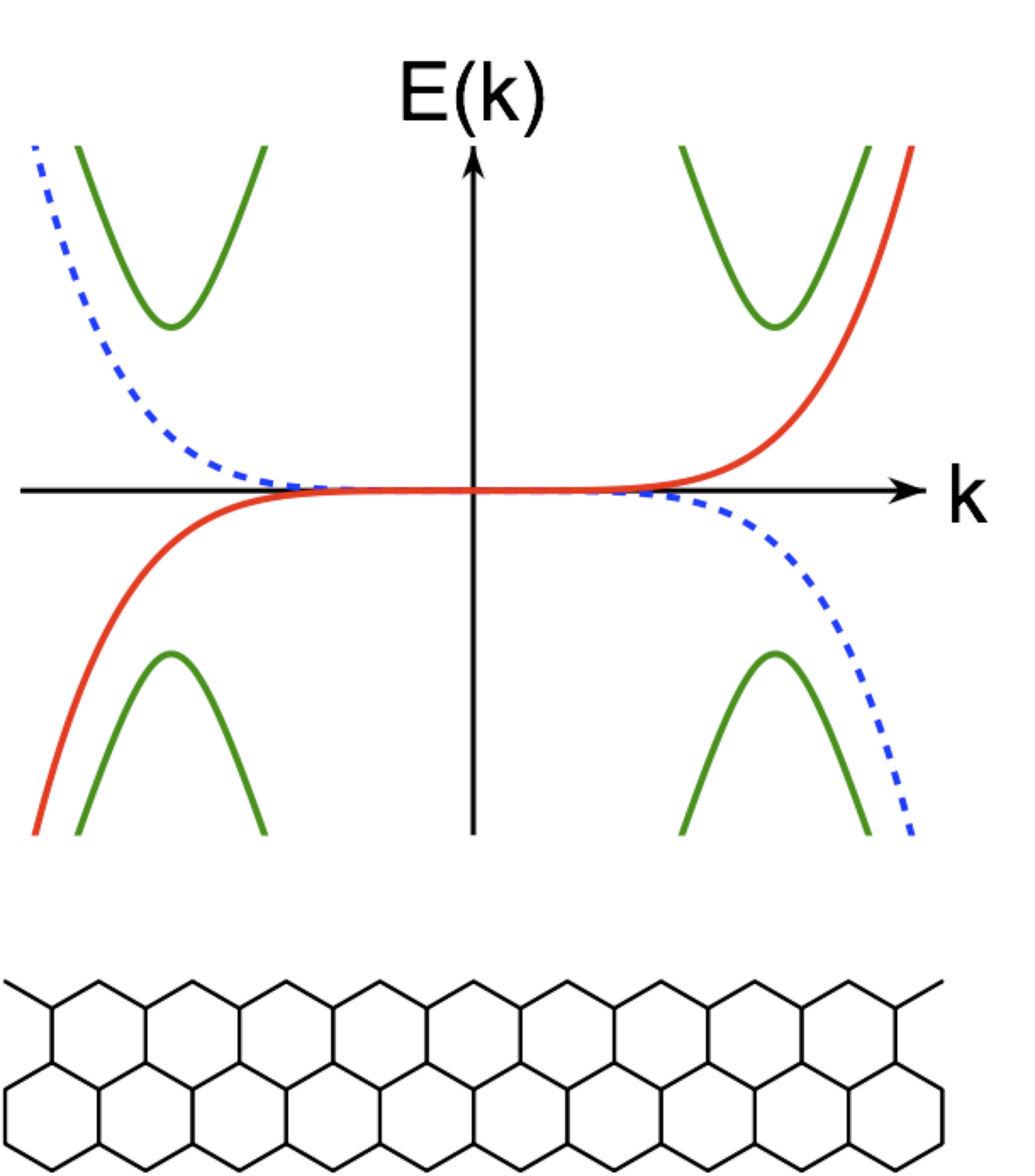}} 
		\caption{Three classes of metallic nanoribbons included in our study and their low-energy band structure. Dashed and solid lines for the lowest bands indicate states with the same symmetry. Armchair ribbons have a linear dispersion relation and their low-energy charge carriers are chiral fermions. The low-energy charge carriers in zigzag and anti-zigzag ribbons are exponentially localized edge states. Zigzag ribbons are symmetric under reflections through the ribbon axis. Anti-zigzag ribbons lack this symmetry.}
		\label{fig:ribbons}
	\end{center}
\end{figure}

We simulated transport in metallic graphene nanoribbons with a rotated gate potential using a non-equilibrium Green function method.\cite{datta1997etm} (See Fig.~\ref{fig:device}.) To compare the conductance properties of chiral fermions and edge states, we analyzed ribbons with three different edge geometries: armchair, zigzag, and antizigzag. (See Fig.~\ref{fig:ribbons}.) Rotating the gate potential allowed us to investigate the effects of band mixing as well.

\begin{figure}[htb]
		\begin{center}
	\subfigure[Armchair Ribbon]{\includegraphics[width=0.45\textwidth]{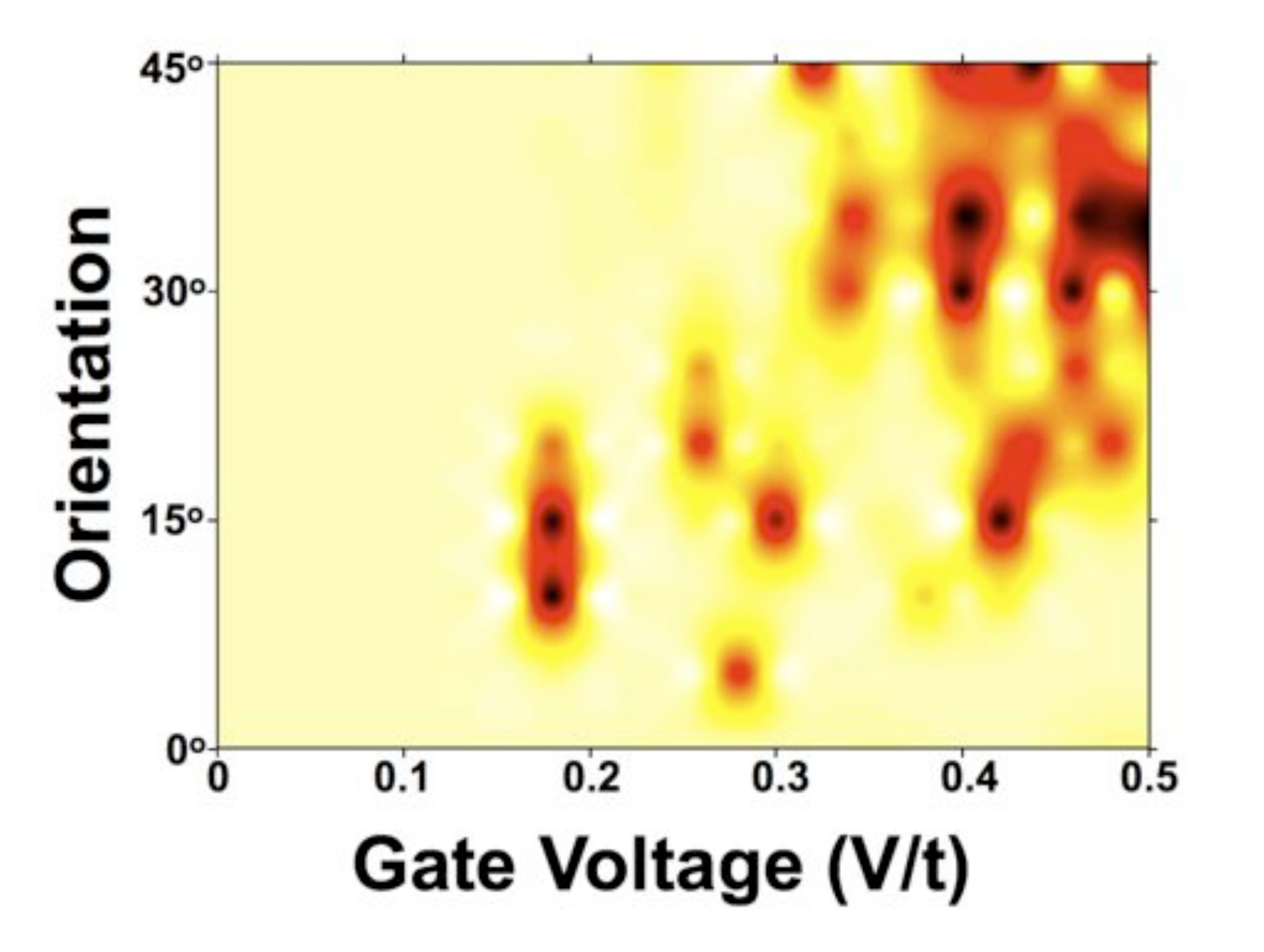}} \\
	\subfigure[Zigzag Ribbon]{\includegraphics[width=0.45\textwidth]{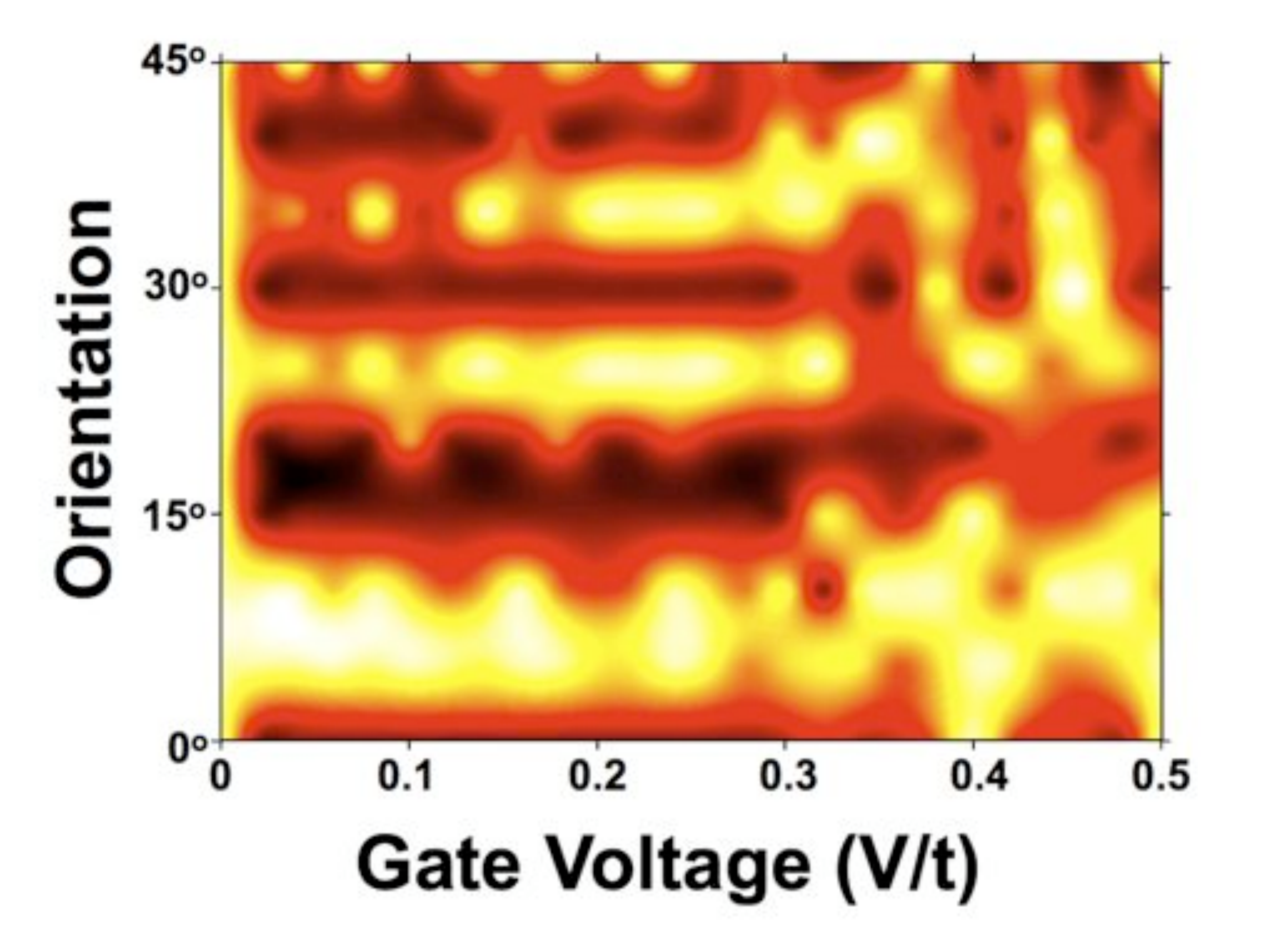}}
	\subfigure[Anti-zigzag Ribbon]{\includegraphics[width=0.45\textwidth]{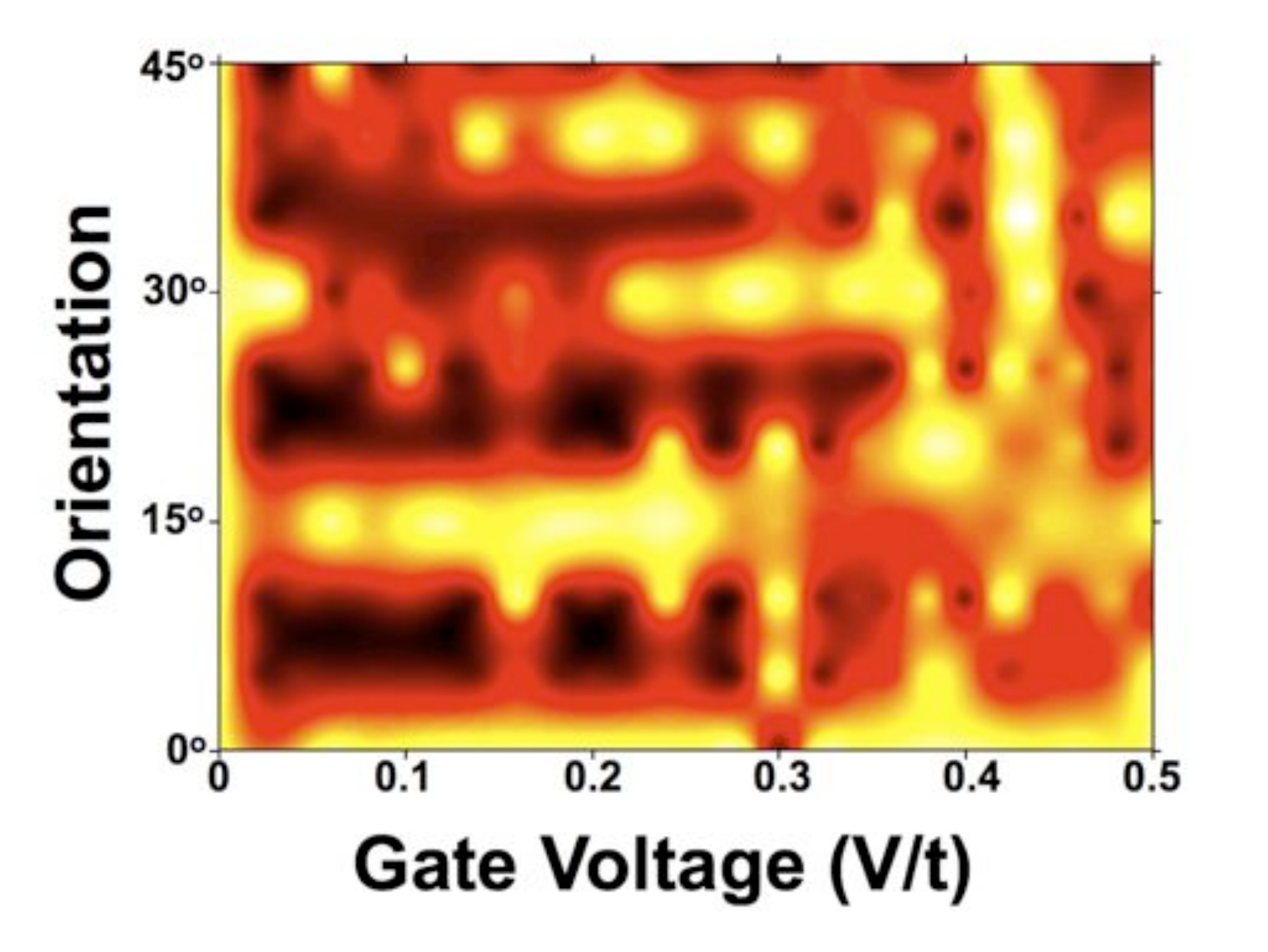}}
		\caption{Transmission probability as a function of gate voltage and orientation. Bright regions indicate a transmission probability near 1 and dark regions indicate a transmission probability near 0. For these data, the fermi energy was $0.01 \, t \approx 30$ meV, and the gate width was 5 nm.}
		 \label{fig:transmission}
	\end{center}
\end{figure}

Fig.~\ref{fig:transmission} shows the transmission probability $T(V_G,\phi)$ as a function of gate voltage and orientation for all three edge geometries. Fig.~\ref{fig:transmission}(a) illustrates our primary result: the perfect reflection of chiral fermions. Ribbons with armchair edges exhibit \emph{resonant backscattering}: nearly perfect transmission for most gate configurations, and nearly perfect reflection near isolated combinations of gate voltage and orientation.

In the remainder of this Letter, we describe our numerical method and results in more detail, discuss the origin of resonant backscattering, and show that the crossover from perfect transmission to perfect reflection should be observable in graphene constrictions gated by a charged carbon nanotube.

The core of our study was a series of numerical transport simulations based on the non-equilibrium Green function (NEGF) method.\cite{datta1997etm} We calculated transmission through infinite graphene ribbons with rotated gate potentials as shown in Fig.~\ref{fig:device}. In our calculations, an infinite ribbon is divided into three sections: two semi-infinite graphene ``leads'' and a graphene ``sample'' that includes the gated region. The probability of transmission between the two leads gives the conductance of the device in units of $e^2/h$. We studied transport at infinitesimal bias, which is equivalent to linear response theory.

The Hamiltonian for the leads and the sample was a $\pi$-electron tight-binding model with nearest-neighbor hopping ($t = 2.7$ eV) on a honeycomb lattice with lattice spacing $a = 0.25$ nm. We approximated the gate by an on-site potential. In the \method{}, the leads introduce a self-energy to the sample Hamiltonian. This self-energy only depends on the surface Green function at the points of contact between the lead and sample, which we calculated with a self-consistent renormalization method.\cite{lopezsancho1985hcs}

Using the model in Fig.~\ref{fig:device}, we studied transport in ribbons with three different edge geometries: armchair, zigzag, and anti-zigzag. (See Fig.~\ref{fig:ribbons}.) For each edge structure, we simulated ribbon widths between 2 and 10 nm. For each ribbon geometry, we varied the gate voltage, width, and orientation. We also analyzed three different profiles of the gate potential: a square barrier, a truncated parabolic barrier, and a gaussian barrier.

Fig.~\ref{fig:transmission} is representative of our general results. It shows the transmission probability $T(V_G,\phi)$ for a 5 nm square barrier as a function of gate voltage and orientation for ribbons with different edge structures, but similar widths: armchair (4.9 nm), zigzag (3.5 nm), and anti-zigzag (3.3 nm). The fermi level was $0.01 t$, or about 27 meV. White regions indicate perfect transmission and black regions indicate perfect reflection. Each plot combines data from 260 gate configurations using MATLAB's biharmonic spline interpolation function.

\textbf{Armchair ribbons:}  Fig.~\ref{fig:transmission}(a) shows that transmission is nearly perfect for most gate configurations, but there is nearly perfect reflection near a few isolated voltage-angle combinations. For example, the transmission probability at $V_G = 0.18 t$ and $\phi = 10^\circ$ is $T \approx 4 \cdot 10^{-4}$.

Since backscattering of chiral fermions is ``forbidden,'' the unusual feature of Fig.~\ref{fig:transmission}(a) is not the large regions of nearly perfect transmission; it is the small regions of nearly perfect reflection.

A scalar potential cannot directly couple chiral fermions moving in opposite directions; however, it can couple states in different bands. Electrons are confined along the width of the ribbon, and the different subbands are analogous to the modes of a particle in a box. Any potential that is nonuniform along the width of the ribbon will couple these modes. Except when $\phi = 0$, the rotated gate potential couples states in different bands. Reflection occurs by way of an indirect coupling mediated by virtual transitions within the gated region.

Scattering theory indicates that the splitting of nearly degenerate bands by trigonal warping is necessary for resonant backscattering. The first-order Born approximation for the reflection amplitude vanishes since it describes reflection of chiral fermions by a scalar potential. In a three-band model like that depicted in Fig.~\ref{fig:ribbons}(a), the second-order Born approximation is proportional to the splitting of the two neighboring bands. If these bands are degenerate (as predicted by the effective Dirac theory), the second- and third-order terms of the Born series vanish, and it seems likely all terms in the series will vanish.

Perturbation theory shows that reflection of chiral fermions by a scalar potential is possible, but it does not accurately describe the numerical data. Further analysis is necessary to determine the mathematical form of the resonance conditions.

\textbf{Zigzag Ribbons:} Figs.~\ref{fig:transmission}(b) and (c) illustrate the differences between chiral fermions and transport by edge states. The conductance oscillates as the gate is rotated, and transmission in zigzag and anti-zigzag ribbons is complementary: conditions that maximize the transmission in one ribbon minimize it in the other.

Recent theoretical studies of transport in zigzag ribbons found similar transmission characteristics. Li \emph{et al.}~reported complementary transmission probabilities through gated regions with $\phi=0$ and finite width,\cite{li2008rst} and Akhmerov \emph{et al.}~reported oscillations and complementary transmission probabilities in semi-infinite $pn$ junctions with arbitrary orientation.\cite{akhmerov2008tvv} Fig.~\ref{fig:transmission} shows that these features also arise in $pnp$ junctions of finite width, and are probably due to scattering at the interface between regions with different potentials, not within the gated region.

The transmission probability in zigzag and antizigzag ribbons oscillates with the same period, but out of phase by $\pi/2$.\cite{akhmerov2008tvv} The maximum or minimum transmission probability is expected when $\tan \phi = 3 N a / R$, where $a$ is the lattice spacing, $R$ is the ribbon width, and $N$ is an integer. For the ribbons in Fig.~\ref{fig:transmission}, this expression predicts angles of $12^\circ$, $23^\circ$, and $33^\circ$, in good agreement with the data.

As the gate voltage is increased, virtual transitions to other bands within the gated region disrupt the oscillations in Fig.~\ref{fig:transmission}.

Some results of our study not apparent in Fig.~\ref{fig:transmission} inlcude the following.

\emph{Potential Profile}: Changing the potential profile does not affect the qualitative features of the transmission plots in Fig.~\ref{fig:transmission}. We investigated three different potential profiles, holding the width and integrated potential strength fixed: square, truncated parabola, and gaussian. In armchair ribbons, resonant backscattering occurred in all cases, although the location and number of resonances changed with the profile. In zigzag ribbons, all three profiles led to transmission oscillations with maxima and minima at the same angles. The effect of changing the profile is strongest in all ribbons for large gate voltages ($V_G > 0.3 t$), when band mixing effects are important.

The square barrier changes discontinuously on the scale of the lattice spacing. The truncated parabolic barrier is continuous, but its first derivative is discontinuous. The gaussian barrier is smooth on the scale of the lattice spacing. Since all three potential profiles have similar transmission characteristics, resonant backscattering and transmission oscillations are \emph{not} due to abrupt changes in the potential on the scale of the lattice spacing. Rather, they are due to scattering processes that only depend on the symmetry of the potential.

\emph{Ribbon Width}: Changing the width of the ribbon shifts the positions of resonant backscattering peaks in armchair ribbons and the spacing of transmission extrema in zigzag ribbons. Since resonant backscattering involves virtual transitions to neighboring bands, it is sensitive to the band spacing, which is inversely proportional to the width of the ribbon. As discussed above, the period of transmission oscillations in zigzag ribbons is inversely proportional to the ribbon width.

\emph{Gate Width}: Changing the width of the gated region affected the position and number of resonant backscattering peaks in armchair ribbons, but had little effect on transmission in zigzag ribbons. This width dependence suggests that scattering of chiral fermions occurs within the gated region, while edge states are scattered at the boundary of the gated region.

Three questions will determine whether the transmission characteristics in Fig.~\ref{fig:transmission} can be detected in an experimental device. Is it possible to make carbon nanoribbons with well-defined edge structures? Is it possible to generate a gate potential with a well-defined orientation and uniform width on the atomic scale? Will disorder in experimental devices eliminate the interesting effects?

Resonant backscattering and conductance oscillations are due to unusual properties of electrons in the lowest bands of metallic \gnr{}s. To isolate these effects, the band spacing must be larger than the temperature at which measurements are taken. Since the band spacing is roughly 500 $\text{meV}/R$ where $R$ is the width of the ribbon in nanometers, the ribbon width should be less than 20 nm for room-temperature measurements. In addition, resonant backscattering requires the splitting of nearly degenerate bands by trigonal warping, which is largest in narrow ribbons.

Narrow \gnr{} devices might be etched from graphene flakes using iron catalysts\cite{datta2008cef} or an electron microscope.\cite{fischbein2008ebn} Current techniques may not offer the precision necessary to make a long ribbon of uniform width. However, it might be possible to create narrow constrictions with well-defined edges persisting over several nanometers in a large graphene flake. A rotated gate potential could then be applied to the constricted region.

Constructing a gate of with a precise width and orientation may be difficult. One possibility is to use a carbon nanotube, a nanowire, or another \gnr{} as the gate. Charging the nanowire could generate potentials similar to the model system in Fig.~\ref{fig:device}.

A carbon nanotube can be charged to accommodate a few electrons per nanometer. If the nanotube is approximated by a line of charge separated from the \gnr{} by the nanotube radius, the potential due to a charged nanotube with a radius of 1 nm is roughly equivalent to that of a square gate in our model with a voltage of 1 V and a width of 2 nm. Different approximation schemes will result in different values for these parameters, but this estimate suggests that charged carbon nanotubes could generate the potentials required to observe the crossover from perfect transmission to perfect reflection in gated armchair \gnr{}s.

Given a ribbon with a well-defined edge and a gate of uniform width, there is still the question of whether conductance oscillations or resonant backscattering can be observed in the disordered environment of a typical graphene experiment. The chiral fermions in metallic armchair ribbons should be insensitive to this type of disorder. As long as the gate potential provides a coupling between different subbands in the ribbon, it should be possible to directly observe resonant backscattering as a sharp increase in resistance for specific gate configurations. The edge states in zigzag ribbons are scattered by any potential that breaks inversion symmetry, so transport in these ribbons will be sensitive to disorder. Conductance oscillations may be difficult to observe directly.

In summary, the conductance of gated graphene nanoribbons depends strongly on the edge structure of the ribbon. Armchair ribbons exhibit resonant backscattering: perfect transmission except near isolated combinations of gate voltage and orientation where the transmission drops to zero. The effect is equivalent to perfect reflection of chiral fermions. In contrast, the edge states in zigzag ribbons give rise to an oscillating conductance as the gate is rotated.

The chiral fermions in metallic armchair \gnr{}s should be insensitive to minor variations in the potential, making it possible to detect resonant backscattering in constricted regions of graphene flakes gated by a charged carbon nanotube or nanowire. Such a device might eventually be used as a switch in nanoscale electronic devices patterned from graphene.

J.M.K. would like to thank Mike Fischbein, Jens Bardarson, and Paul McEuen for their helpful suggestions.

This work was supported by the Cornell Center for Materials Research.

\end{document}